**Predictive Analysis of *Gmelina arborea* (Melina) Growth in Plantations of Esmeraldas: A Perspective for Silvicultural Management in Tropical Ecuador**

corresponding autor
MSc. José Gabriel Carvajal Benavides [1]
jgcarvajalb@utn.edu.ec
https://orcid.org/0000-0001-9920-4991
Faculty of Engineering in Agricultural and Environmental Sciences.
Forestry Engineering Program
Technical University of the North.

MSc. Hugo Orlando Paredes Rodríguez [2]
hoparedes@utn.edu.ec
https://orcid.org/0000-0002-5880-1607
Faculty of Engineering in Agricultural and Environmental Sciences.
Forestry Engineering Program
Technical University of the North

MSc. Oscar Armando Rosales Enríquez [3]
oarosales@utn.edu.ec
https://orcid.org/0000-0001-7131-6203
Faculty of Engineering in Agricultural and Environmental Sciences.
Forestry Engineering Program
Technical University of the North

MSc. Eduardo Jaime Chagna Avila [4]
ejchagna@utn.edu.ec
https://orcid.org/0000-0003-2527-4334
Faculty of Engineering in Agricultural and Environmental Sciences.
Forestry Engineering Program
Technical University of the North

Xavier Germán Valencia Valenzuela [5]
xavierger@hotmail.com
https://orcid.org/0000-0002-3209-9581
Independent forestry engineer, in private professional practice
Imbabura Association of Forestry Engineers

Mgs. Guillermo David Varela Jácome [6]
gdvarelaj@utn.edu.ec
https://orcid.org/0000-0002-4070-8201
Faculty of Engineering in Agricultural and Environmental Sciences.
Forestry Engineering Program
Technical University of the North

## ABSTRACT

This study presents a rigorous assessment of the growth performance of *Gmelina arborea* (melina) in a 67-hectare plantation located in Chontaduro, Tabiazo Parish, Esmeraldas, Ecuador. The plantation was established in 2017 under a high-density planting system (650 trees/ha). Permanent monitoring techniques were applied in 16 one-hectare plots to analyze structural growth variables, including survival rate, diameter at breast height (DBH), total height, commercial height, basal area, volume, and mean annual increment (MAI).

The results show an average survival rate of 80.2%, with a mean DBH of 25.3 cm at five years, indicating sustained growth under favorable edaphoclimatic conditions. Volume was calculated using the equation $V = G \times HT \times Ff$, yielding average values of 183.262 $m^3$ for total volume and 166.19 $m^3$ for commercial volume. The estimated MAI for diameter and height was 5.06 cm/year and 3.61 m/year, respectively, with values comparable to studies conducted in other Ecuadorian sites, although lower productivity was observed in Esmeraldas, attributed to edaphic and climatic differences identified through soil type and environmental condition analyses.

The research highlights the significant influence of edaphic conditions, silvicultural management, and environmental variables on the performance of *Gmelina arborea* in tropical Ecuador. The findings provide a foundation for optimizing forest management strategies and improving growth indicators in commercial plantations, contributing to the sustainable development of forest resources in the region and strengthening silvicultural planning based on predictive models tailored to local conditions. This study represents a step forward in the scientific assessment of melina growth under Ecuadorian conditions, promoting more precise and sustainable silvicultural practices

## INTRODUCTION

Reforestation in Ecuador has experienced significant growth over the past decades, driven by both public policies and private initiatives (FAO, 2020). According to the Ministry of Agriculture, Livestock, Aquaculture and Fisheries (MAGAP, 2015), the forest species with the highest demand for implementation of projects under the Forest Incentives Program (PIF) include teak (*Tectona grandis*) with 12,561.06 ha, *Gmelina arborea* (melina) with 5,294.57 ha, balsa (*Ochroma pyramidale*) with 6,274.32 ha, patula pine (*Pinus patula*) with 1,604.88 ha, radiata pine (*Pinus radiata*) with 2,017.48 ha, chuncho (*Cedrelinga cateniformis*) with 664.77 ha, and blue gum (*Eucalyptus globulus*) with 627.77 ha.

Currently, according to the Angiosperm Phylogeny Group (APG) classification, *Gmelina arborea* Roxb. ex Sm. belongs to the family Lamiaceae, although in literature prior to 2016 it was classified within Verbenaceae (GBIF, 2023). This exotic species in Ecuador originates from Southeast Asia, particularly from the sub-Himalayan region, and has successfully spread in tropical regions due to its remarkable adaptability (Chazdon, 2014).

*Gmelina arborea* Roxb. ex Sm. is considered an exotic species in Ecuador as it is native to Asia, especially India in the sub-Himalayan region, and is sporadically found in western and southern India. It has been introduced to countries within the tropical belt; it is a tree reasonably strong for its weight (Ecuador Forestal, 2023).

Noted for its rapid growth, *Gmelina arborea* has been adopted in numerous developing countries as an alternative for short- and medium-term timber production (Paquette & Messier, 2010). Its optimal establishment occurs between 0 and 900 meters above sea level, under annual precipitation ranging from 1,000 to 4,000 mm, in life zones that include tropical dry forest, tropical moist forest, and tropical very moist forest, preferably on soils with loam to clay-loam textures (Moya & Muñoz, 2010).

Nevertheless, its intolerance to strong winds and waterlogged soils necessitates adaptive silvicultural management that considers local edaphic characteristics to ensure its success (Brockerhoff et al., 2013). In Ecuador, particularly in Esmeraldas, these variables represent a challenge that underscores the importance of predictive research aimed at optimizing sustainable forest productivity.

## METHODOLOGY

The Melina plantation is located in the area known as Chontaduro, in the parish of Tabiazo, Esmeraldas canton and province. The property, owned by Forest Engineer Marcelo Estévez, has a total area of 152.80 ha.

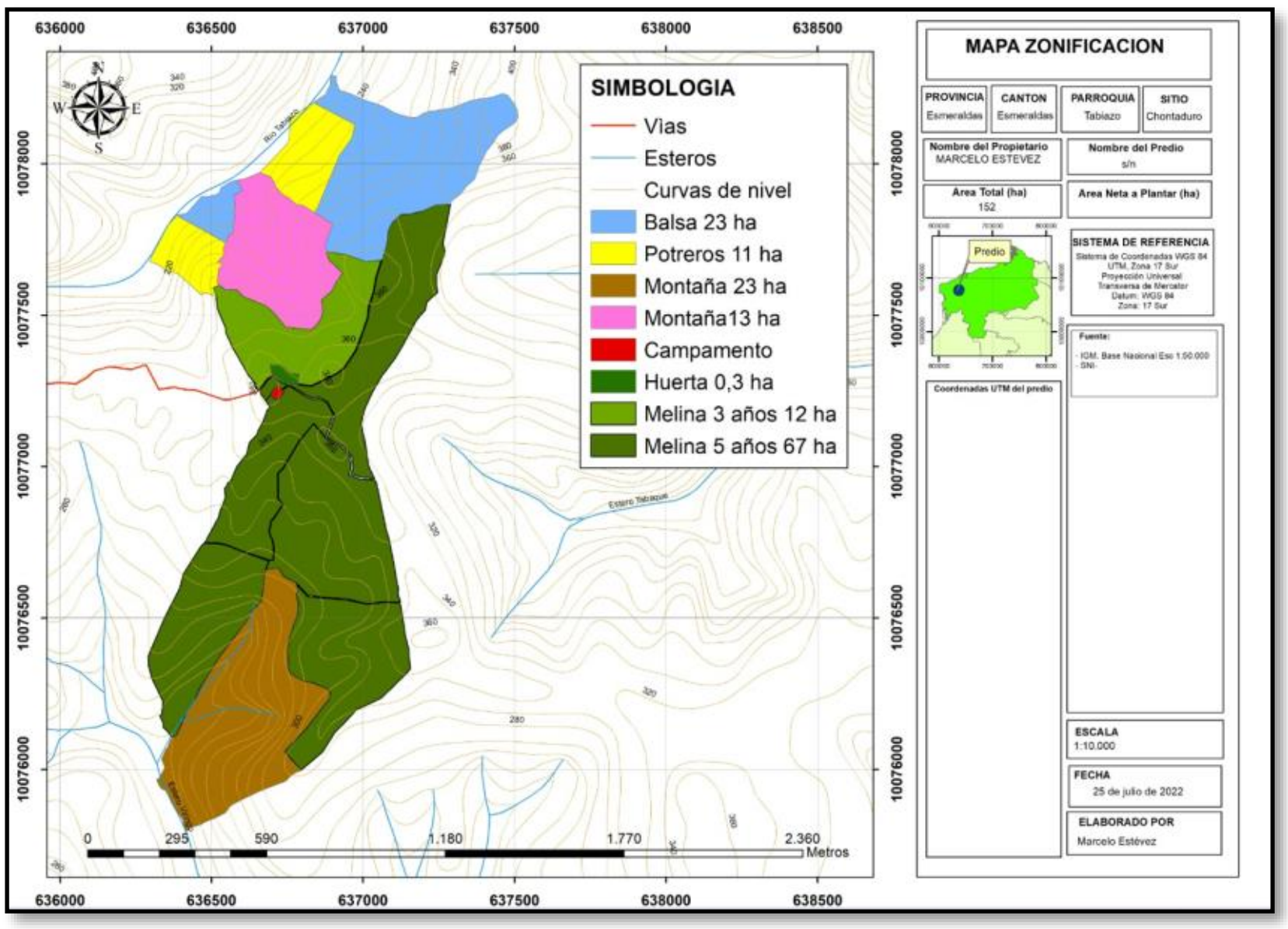

Figure 1. Political location of the plantation.

According to the life zone classification established by Holdridge in 1978, the area corresponds to a Tropical Dry Megathermal zone. The altitudinal range varies between 26 and 365 meters above sea level. The climate is characterized by an average temperature of 26 °C, with average maximum and minimum temperatures of 30 °C and 21 °C, respectively, and an average annual precipitation of 1,500 mm, with the highest rainfall occurring from January to July. The soils are loam-clay, loam-sandy, and stony, and are moderately deep (PDOT TABIAZO, 2015).

**Evaluated Plantation**

The plantation was established in April 2017, with a net planted area of 67 hectares. The genetic material used originated from seeds imported from CATIE (Costa Rica), which were used for propagation and seedling production in polyethylene bags.

Land preparation was performed manually, with the area cleared and readied for planting. The planting spacing was 4 x 3.85 m, with an initial density of 650 trees per hectare. During tree growth, manual weeding was performed three times per year during the first two years. Due to the species' rapid growth, silvicultural activities began in the first year after establishment, with formative pruning continuing for the following five years to obtain knot-free timber, pruning up to a height of 4.5 meters.

**Sample Size**

The established population was 67 ha (N=67), from which a sampling intensity of 24% was calculated (n=16 sample plots of 1 ha each), using a regular design of 100 x 100 m (10,000 $m^2$). The plots were systematically distributed according to the planted area (Figure 2), based on the guidelines of the National Financial Corporation (CFN).

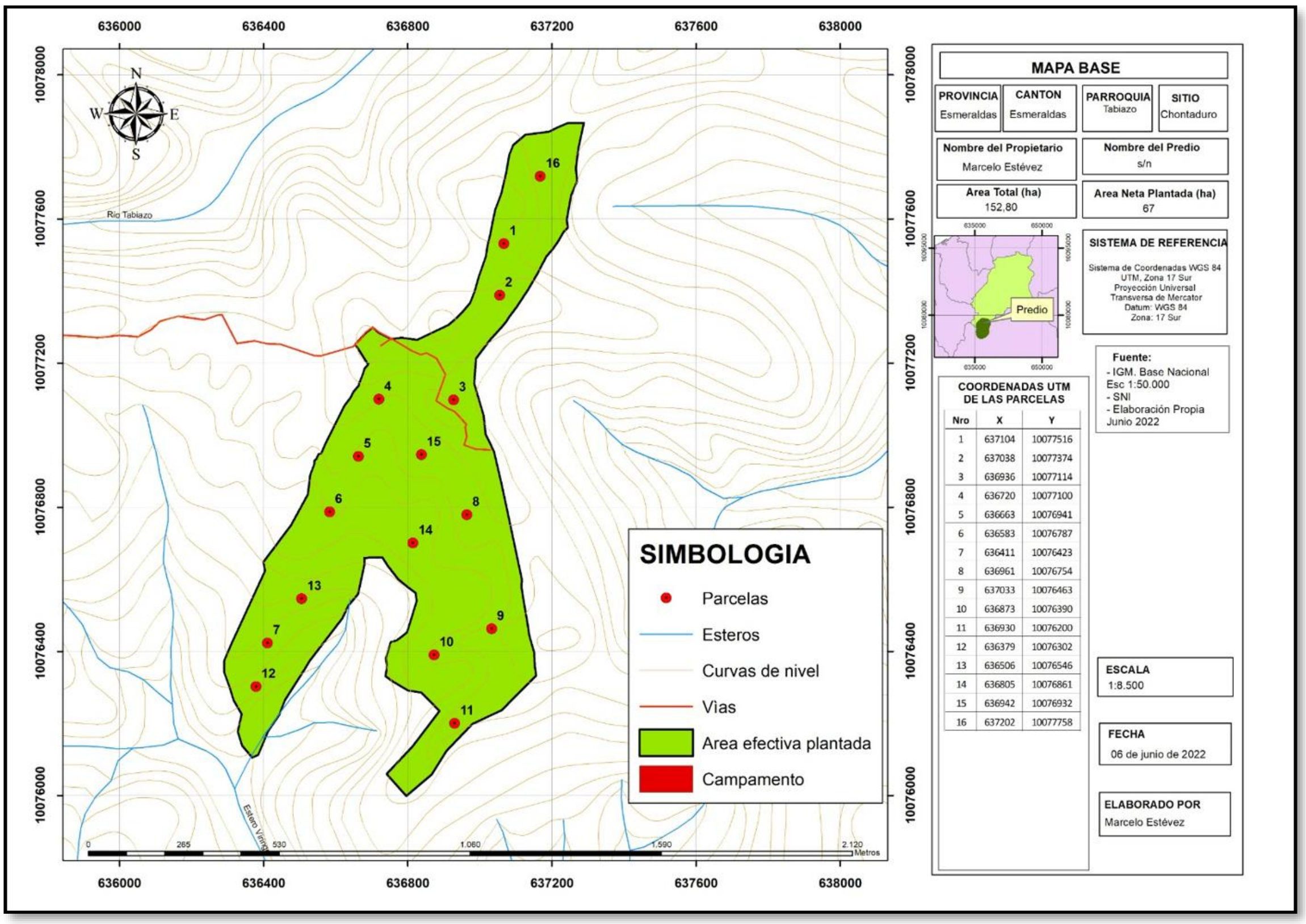

Figure 2. Location of the plantation and sample plots.

**Growth and Productivity Assessment of the Melina Plantation**

A diameter tape was used to measure the diameter at breast height (DBH) at 1.3 m above the base. Total and commercial heights (m) were estimated using a Suunto hypsometer. Productivity was assessed based on the following indirect variables: first-year survival, mean annual increment (MAI) in DBH, total height, basal area, and volume. Survival was determined using the following equation:

According to Linares (2005), and as modified by the author, the formula to determine survival is:

%Sv=Number of live plantsTotal /number of established plants×100 (Eq.1)

%Sv=Total number of established / plantsNumber of live plants×100 (Eq.1)

To calculate basal area, Equation (2) was used:

Basal area (m2):$G=\pi 4\times(DBH)\ 2$ (Eq.2)

To calculate volume, Equation (3) was used:

$V=G\times HT\times Ff(Eq.3)V=G\times HT\times Ff$ (Eq.3)

Where:

V = volume in $m^3$

G = basal area in $m^2$

Ff = form factor (0.7), according to Ministerial Agreement 095 MAG

MAI (Mean Annual Increment):

The mean annual increment value expresses the average total growth at a given tree age. It represents the average annual growth at any age. The mean annual increment was calculated for diameter, height, basal area, and volume using the following formula:

$IMA=Ytt0(Eq.4)IMA=t0Yt$ (Eq.4)

Where:

IMA = tree volume / age

t0$t$0 = age from time zero

Y = value of the variable considered

## RESULTS

### Growth Variables of the Melina Plantation

The average first-year survival rate was 80.2%, with variation between 36.9% and 95.4% across the 16 plots. The mean diameter at breast height (DBH) at five years after establishment was 25.3 cm, increasing from 4.6, 10.2, 16.1, and 19.4 cm at each measurement period (Table 1).

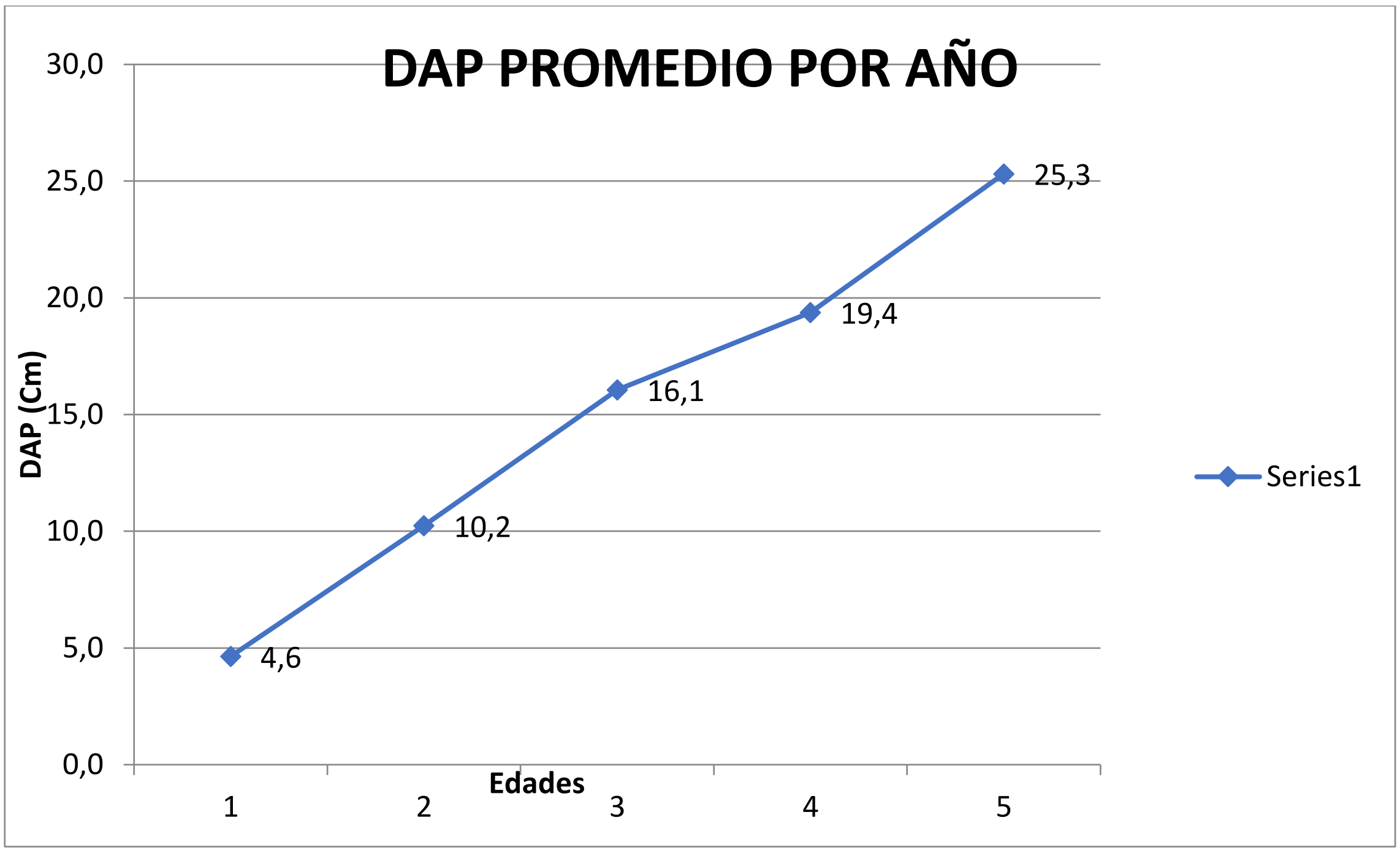

Illustration 1. Mean annual DBH growth.

Table 1. Summary of Dasometric Variables for the Melina Plantation

| PARCELA | SOBREVIVENCIA % | ARB TOT | DAP | HT | HC | HPALET | AB | VOL TOTAL | VOL COM | VOL PALET |
|---|---|---|---|---|---|---|---|---|---|---|
| **1** | 73,8 | 330 | 0,24 | 17,78 | 16,13 | 8,8 | 14,72 | 183,262 | 166,192 | 91,148 |
| **2** | 81,5 | 350 | 0,24 | 19,00 | 16,46 | 7,5 | 15,46 | 205,653 | 178,130 | 81,643 |
| **3** | 84,6 | 340 | 0,24 | 20,53 | 18,76 | 7,5 | 16,02 | 230,233 | 210,442 | 84,111 |
| **4** | 95,4 | 350 | 0,25 | 18,11 | 15,71 | 6,4 | 16,88 | 214,079 | 185,715 | 75,299 |
| **5** | 95,4 | 400 | 0,25 | 19,10 | 16,85 | 7,9 | 19,20 | 256,769 | 226,492 | 106,203 |
| **6** | 95,4 | 460 | 0,25 | 18,37 | 16,85 | 7,5 | 23,19 | 298,181 | 273,480 | 121,037 |
| **7** | 81,5 | 390 | 0,27 | 18,44 | 16,59 | 8,7 | 22,65 | 292,346 | 263,071 | 137,431 |
| **8** | 76,9 | 300 | 0,25 | 18,03 | 15,63 | 6,2 | 14,87 | 187,733 | 162,748 | 64,891 |
| **9** | 92,3 | 370 | 0,25 | 17,92 | 15,62 | 7,4 | 18,52 | 232,320 | 202,535 | 95,311 |
| **10** | 92,3 | 370 | 0,25 | 17,62 | 14,76 | 6,6 | 18,20 | 224,472 | 187,978 | 83,660 |
| **11** | 87,7 | 310 | 0,27 | 17,10 | 14,45 | 7,4 | 17,63 | 211,000 | 178,354 | 91,566 |
| **12** | 93,8 | 400 | 0,25 | 18,55 | 18,55 | 8,8 | 20,17 | 261,889 | 261,889 | 124,591 |
| **13** | 83,1 | 340 | 0,25 | 14,38 | 16,59 | 7,9 | 17,28 | 173,929 | 200,623 | 94,967 |
| **14** | 50,8 | 330 | 0,25 | 17,30 | 16,15 | 7,2 | 16,20 | 196,203 | 183,145 | 81,780 |
| **15** | 36,9 | 240 | 0,27 | 18,13 | 15,50 | 6,2 | 14,04 | 178,185 | 152,379 | 61,033 |
| **16** | 61,5 | 400 | 0,26 | 18,18 | 16,53 | 8,2 | 20,60 | 262,146 | 238,347 | 117,911 |
| **Promedio** | 80.2% | | | | ∑ | Suma | 285,65 | 3608,40 | 3271,52 | 1512,58 |
| | 95% | t studen (16-1) | 2,131 | | $\bar{x}$ | Promedio | 17,85 | 225,52 | 204,47 | 94,54 |
| | | | | | Mx | Máximo | 23,19 | 298,18 | 273,48 | 137,43 |

| | | | | | |
|---|---|---|---|---|---|
| Mi | Mínimo | 14,04 | 173,93 | 152,38 | 61,03 |
| s | DesvT | 2,75 | 39,30 | 37,91 | 21,72 |
| $s^2$ | Varianza | 7,54 | 1544,49 | 1436,92 | 471,76 |
| CV | Coeficiente de variación | 15,4 | 17,4 | 18,5 | 23,0 |
| Sy | Error estándar | 0,71 | 10,15 | 9,79 | 5,61 |
| Em | Error de muestreo | 1,5 | 21,6 | 20,9 | 12,0 |
| Ls | Límite superior | 18,56 | 235,67 | 214,26 | 100,14 |
| Lm | Límite inferior | 17,14 | 215,38 | 194,68 | 88,93 |

Averages and Statistical Summary (n=16):

Mean survival: 80.2%

Mean DBH: 0.25 m

Mean total height: 17.85 m

Mean commercial height: 225.52 m

Mean pallet height: 204.47 m

Mean basal area: 94.54 m²

Mean total volume: 285.65 m³

Mean commercial volume: 3,608.40 m³

Mean pallet volume: 1,512.58 m³

**Productivity of the Melina Plantation**

The mean annual increment (MAI) in DBH was 5.06 $cm \cdot year^{-1}$, with a maximum of 5.46 $cm \cdot year^{-1}$ in plot 15 and a minimum of 4.74 $cm \cdot year^{-1}$ in plot 2. The average MAI for height was 3.61 $m \cdot year^{-1}$, with a maximum of 4.11 $m \cdot year^{-1}$ in plot 3 and a minimum of 2.88 $m \cdot year^{-1}$. The plantation recorded mean MAI values for basal area of 3.57 $m^2 \cdot ha^{-1} \cdot year^{-1}$ and for commercial volume of 45.10 $m^3 \cdot ha^{-1} \cdot year^{-1}$. The highest basal area increment (4.64 $m^2 \cdot ha^{-1} \cdot year^{-1}$) was observed in plot 6, and the lowest (2.81 $m^2 \cdot ha^{-1} \cdot year^{-1}$) in plot 15. Commercial volume and total volume followed the same trend, with maximums of 54.70 and 59.64 $m^3 \cdot ha^{-1} \cdot year^{-1}$, and minimums of 30.48 and 34.79 $m^3 \cdot ha^{-1} \cdot year^{-1}$, respectively.

Table 2. Productivity Summary for Gmelina arborea at Five Years

| PARCELAS | IMA (DAP $cm^{-1}$ año) | IMA (ALTURA $m^{-1}$ .año$^{1}$) | IMA (ÁREA BASAL $m^2$ ha-$^{1}$ .año-$^{1}$) | IMA (VOLUMEN COMERCIAL $m^3$ ha-$^{1}$ .año-$^{1}$) | IMA (VOLUMEN TOTAL $m^3$ ha-$^{1}$ .año-$^{1}$) |
|---|---|---|---|---|---|
| 1 | 4,77 | 3,6 | 2,94 | 33,238 | 36,652 |
| 2 | 4,74 | 3,8 | 3,09 | 35,626 | 41,131 |
| 3 | 4,90 | 4,1 | 3,20 | 42,088 | 46,047 |
| 4 | 4,96 | 3,6 | 3,38 | 37,143 | 42,816 |
| 5 | 4,94 | 3,8 | 3,84 | 45,298 | 51,354 |
| 6 | 5,07 | 3,7 | 4,64 | 54,696 | 59,636 |
| 7 | 5,44 | 3,7 | 4,53 | 52,614 | 58,469 |
| 8 | 5,02 | 3,6 | 2,97 | 32,550 | 37,547 |
| 9 | 5,05 | 3,6 | 3,70 | 40,507 | 46,464 |
| 10 | 5,00 | 3,5 | 3,64 | 37,596 | 44,894 |
| 11 | 5,38 | 3,4 | 3,53 | 35,671 | 42,200 |
| 12 | 5,07 | 3,7 | 4,03 | 52,378 | 52,378 |
| 13 | 5,09 | 2,9 | 3,46 | 40,125 | 34,786 |
| 14 | 5,00 | 3,5 | 3,24 | 36,629 | 39,241 |
| 15 | 5,46 | 3,6 | 2,81 | 30,476 | 35,637 |
| 16 | 5,12 | 3,6 | 4,12 | 47,669 | 52,429 |
| **Promedio** | **5,06** | **3,61** | **3,57** | **40,89** | **45,10** |
| **Max** | **5,46** | **4,11** | **4,64** | **54,70** | **59,64** |
| **Min** | **4,74** | **2,88** | **2,81** | **30,48** | **34,79** |

MAI: Mean Annual Increment.

To determine the best-fit equation, Statgraphics software was used, yielding the following results:

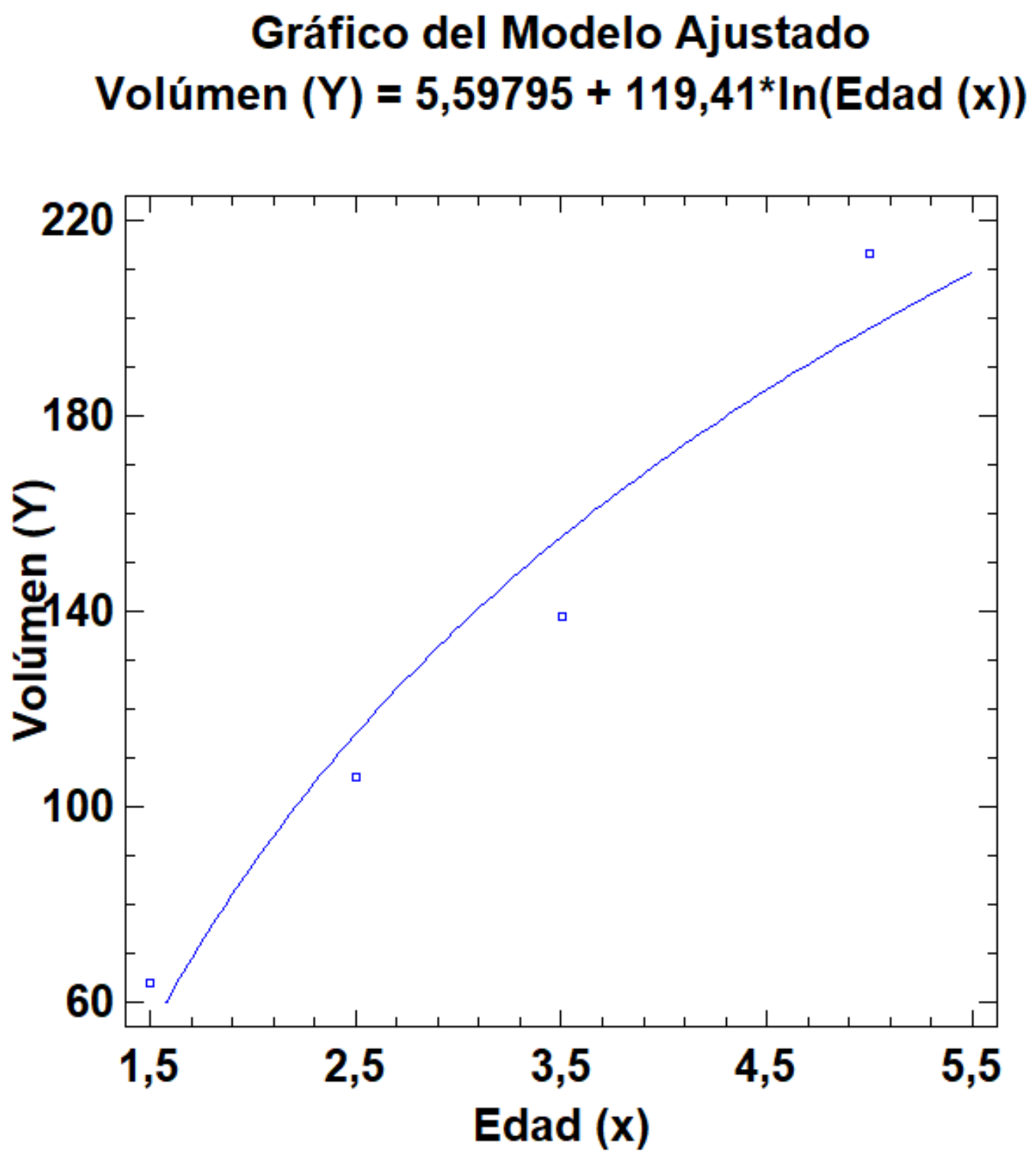

Simple Regression – Volume (Y) vs. Age (x)

Dependent variable: Volume (Y) ($m^3$/ha)

Independent variable: Age (x) (years)

Log-X model: $Y = a + b \cdot \ln(X)$

Adjusted model: Volume (Y) = 5.59795 + 119.41·ln(Age (x))

| Parameter | Estimate | Standard Error | t-Statistic | p-Value |
|---|---|---|---|---|
| Intercept | 5.59795 | 23.5184 | 0.238025 | 0.8340 |
| Slope | 119.41 | 20.6994 | 5.7688 | 0.0288 |

**Comparison of Site Conditions**

The Gmelina plantation site was compared with the climatic and edaphic characteristics of the Esmeraldas Tachina, Santo Domingo, and San Juan La Maná meteorological stations (INAMHI). Ombrothermic diagrams for these stations, based on 25-year. Mean precipitation and temperature data, are shown in:

Figure 3. Average precipitation and temperature at the Esmeraldas-Tachina Station (INAMHI, 2019)

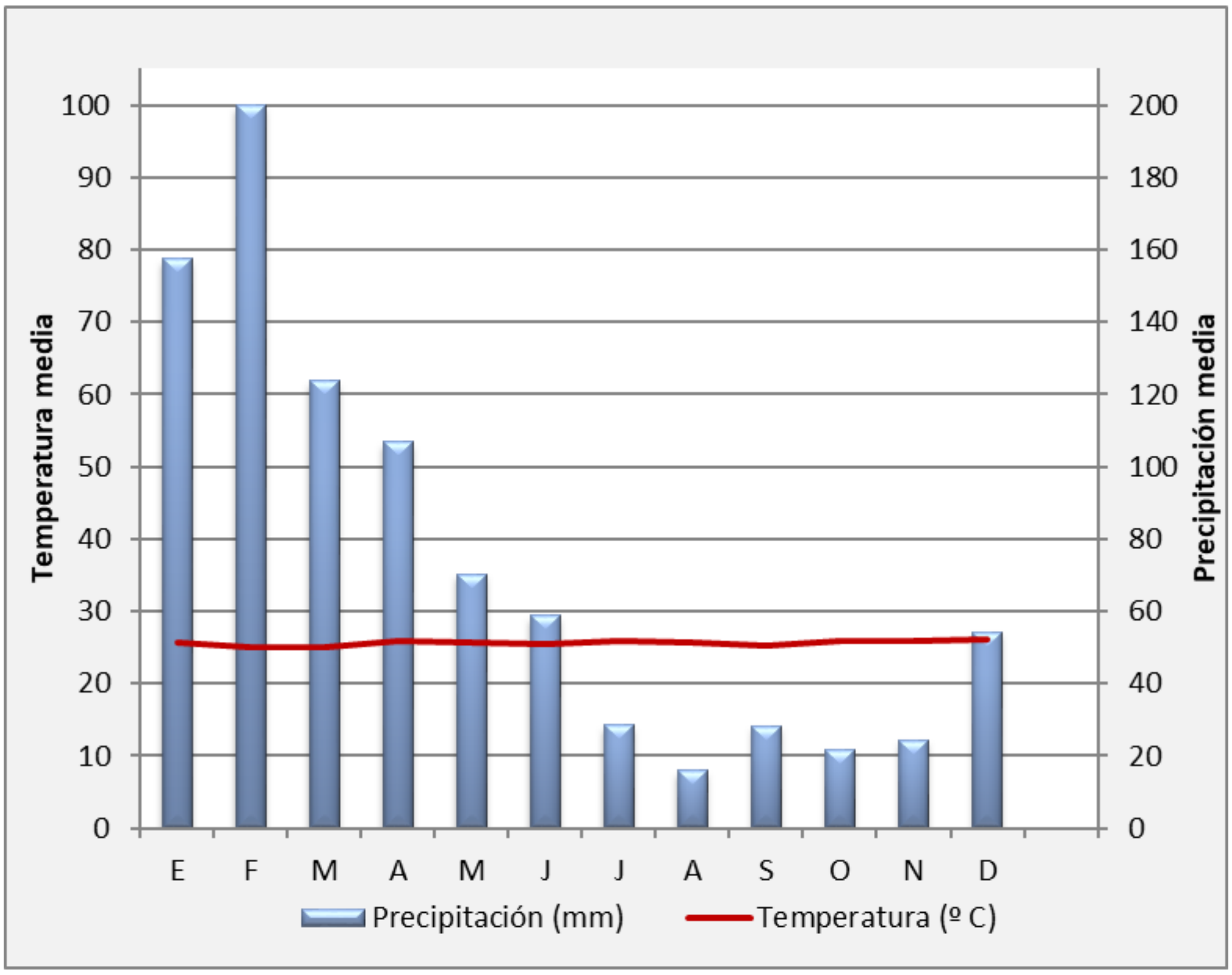

Figure 4. Average precipitation and temperature at the Santo Domingo Station (INAMHI, 2019)

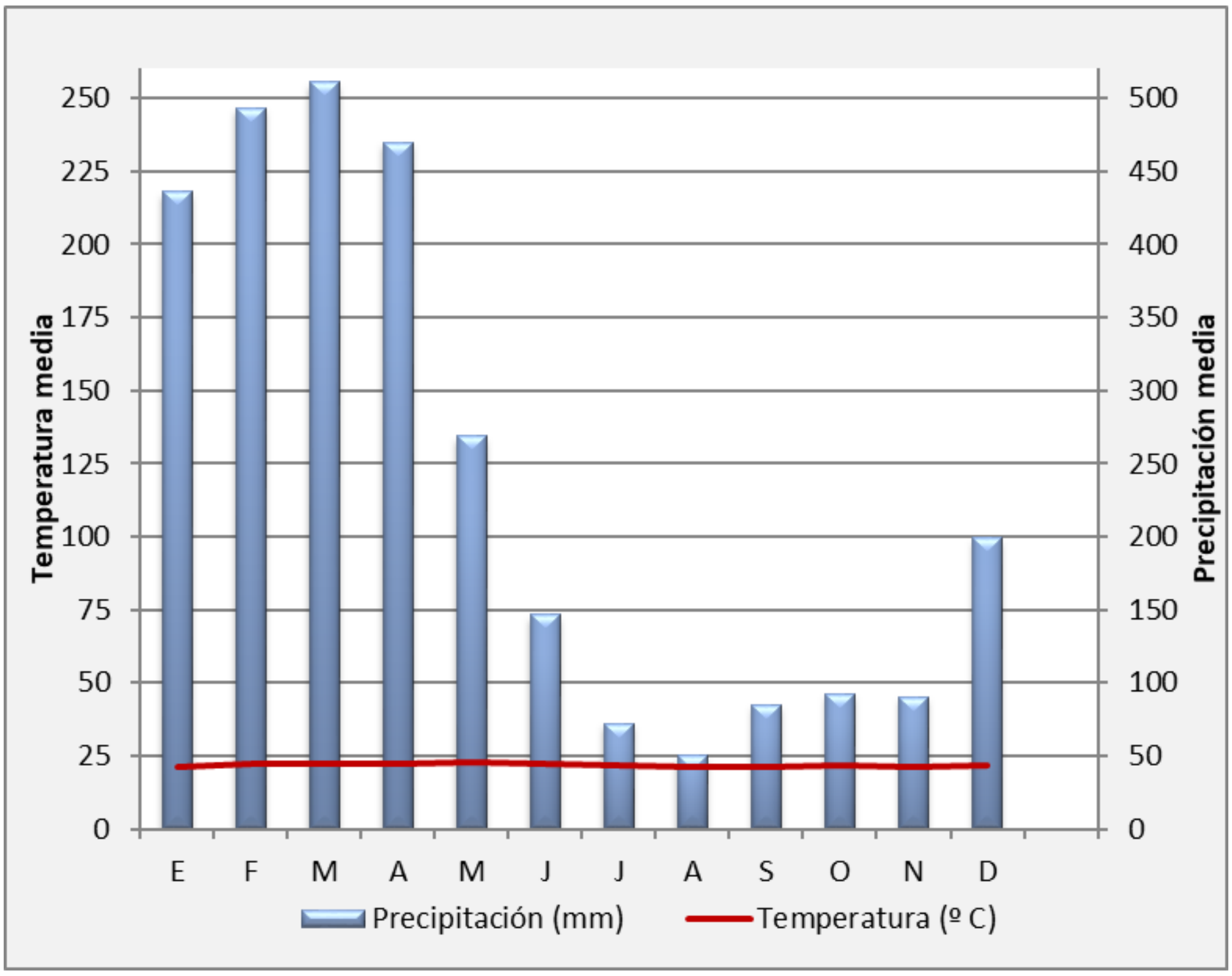

Figure 5. Average precipitation and temperature at the San Juan La Maná Station (INAMHI, 2019).

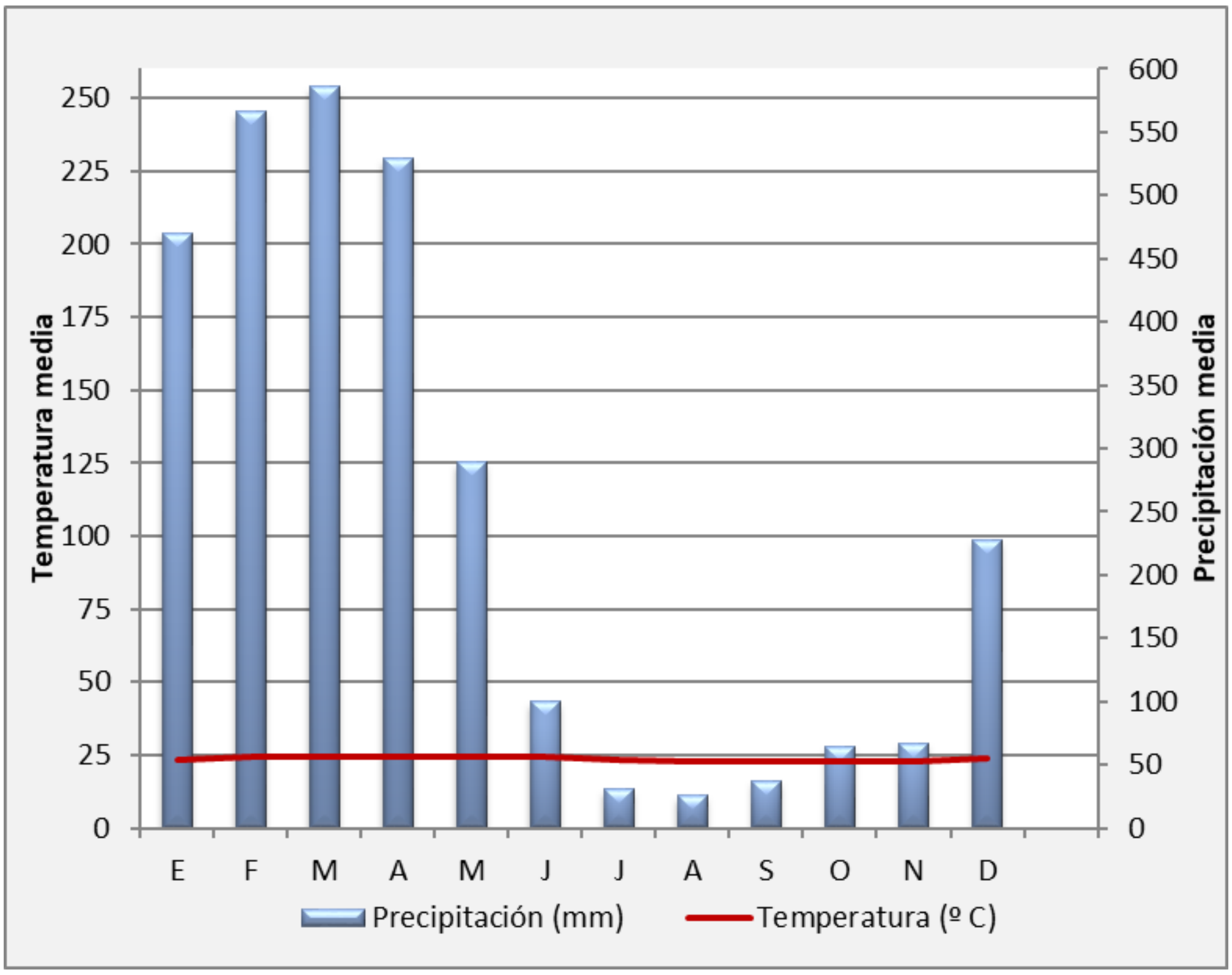

Figures 6 and 7 show the climate types and soil types based on digital cartographic information from the National Information System (SNI).

Figures 6

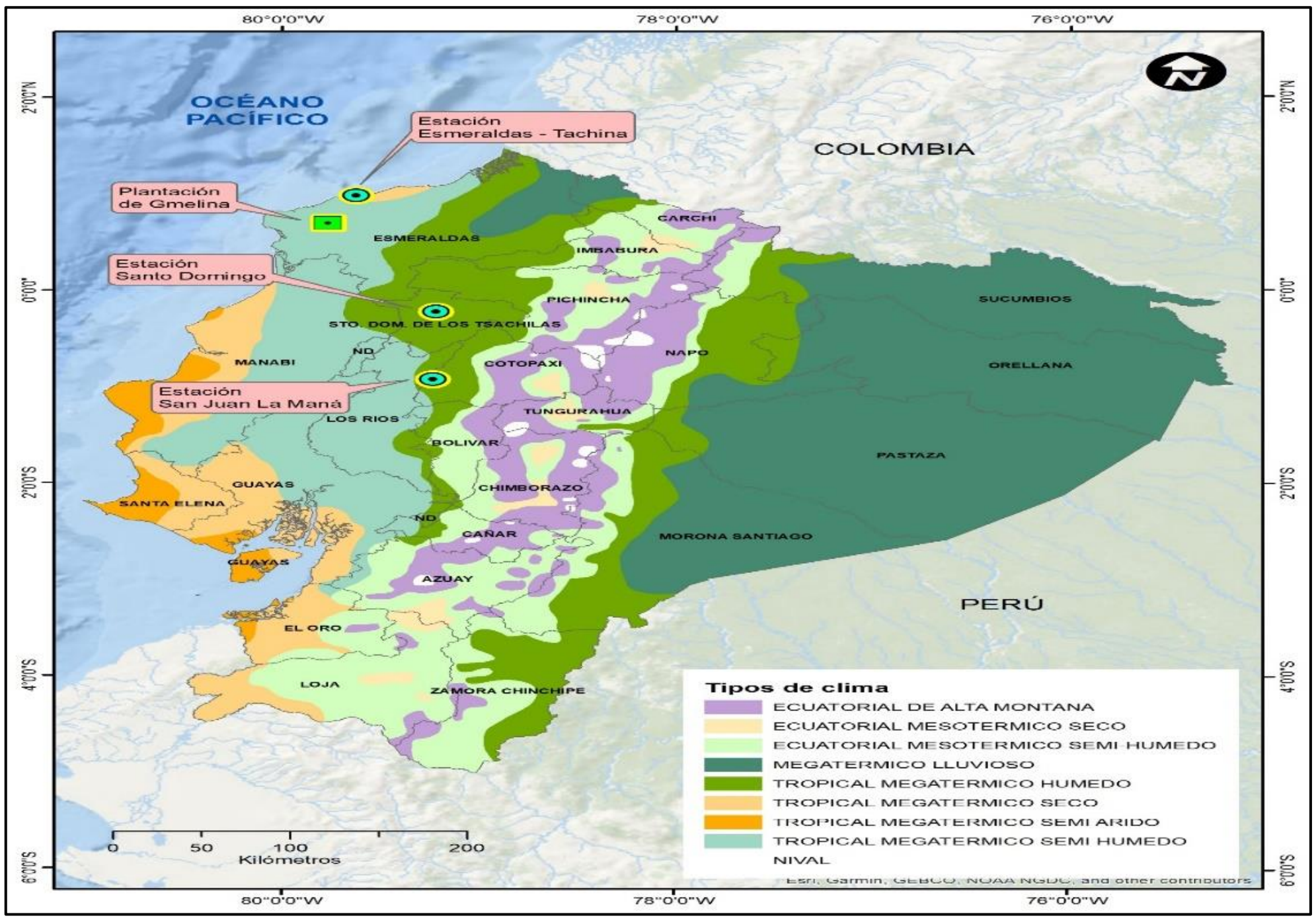

Figure 7 present climate and soil type maps based on digital cartographic information from the National Information System (SNI).

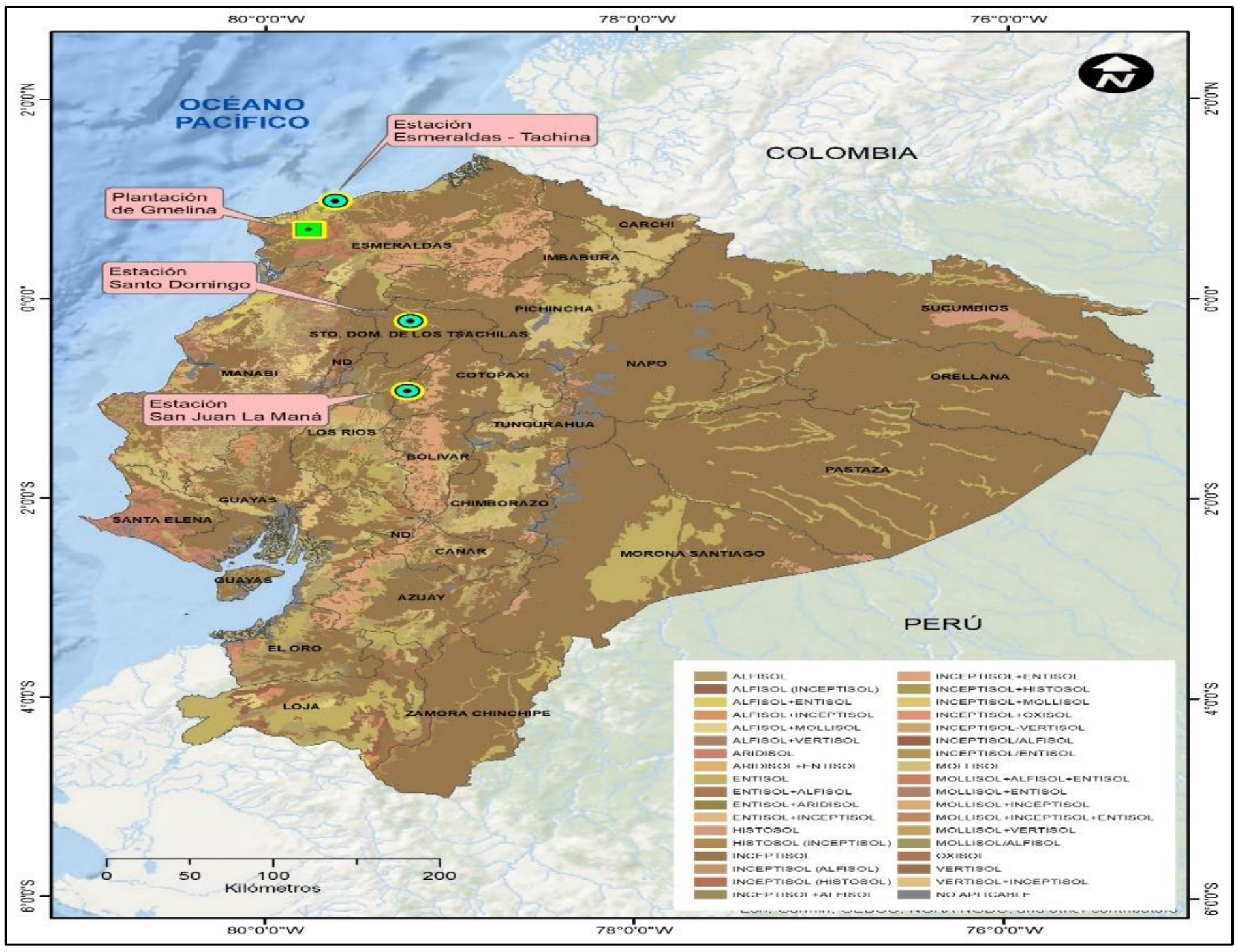

Table 3. Edaphic and Climatic Characteristics Comparison

| Site | Soil Type | Mean Annual Precipitation (mm) | Mean Annual Temperature (°C) | Climate Type | Dry Months |
|---|---|---|---|---|---|
| Gmelina Plantation | Mollisol | 890.1 | 25.6 | Tropical Megathermal Semi-Humid | 5 |
| Santo Domingo | Inceptisol | 2915.4 | 21.9 | Tropical Megathermal Humid | 0 |

| Site | Soil Type | Mean Annual Precipitation (mm) | Mean Annual Temperature (°C) | Climate Type | Dry Months |
|---|---|---|---|---|---|
| La Maná | Inceptisol | 2995.1 | 23.7 | Tropical Megathermal Humid | 3 |

Significant differences exist between the sites, influencing the growth and development of the species. The Gmelina plantation site in this study is markedly different from the others in terms of soil type, mean annual precipitation, mean annual temperature, climate type, and number of ecologically dry months.

**Notes on technical terms:**

DBH: Diameter at Breast Height

HT: Total Height

CH: Commercial Height

BA: Basal Area

MAI: Mean Annual Increment

COMM. VOL: Commercial Volume

PALLET VOL: Volume for pallet use

## DISCUSSION

### Growth Variables of the Melina Plantation

Belgis (2022) reported that the average DBH growth in the Santo Domingo plantation was 22.49 cm $ha^{-1} \cdot year^{-1}$, while in Valencia it was 27.15 cm $ha^{-1} \cdot year^{-1}$ at four years. In Esmeraldas, the average DBH at five years was 25.3 cm $ha^{-1}\cdot year^{-1}$, a value higher than that of Santo Domingo but lower than Valencia; however, it should be noted that there is a one-year difference between studies.

Saltos (2022) found an average of 20.1 cm $ha^{-1}\cdot year^{-1}$ in a five-year-old plantation in Pangua canton, which is lower than the result obtained in this study.

Regarding the mean annual increment (MAI), an average value of 5.06 $cm\cdot year^{-1}$ was obtained. In Belgis (2022), the MAI for Santo Domingo was 5.62 $cm\cdot year^{-1}$ and for Valencia 6.79 $cm\cdot year^{-1}$, values similar to those of Esmeraldas and Santo Domingo.

For average height MAI, the value for Esmeraldas was 3.61 $m\cdot year^{-1}$, lower than those obtained for Santo Domingo (5.91 $m\cdot year^{-1}$) and Valencia (5.52 $m\cdot year^{-1}$).

The commercial volume obtained was 40.89 $m^3$ $ha^{-1}\cdot year^{-1}$, and the total volume was 45.10 $m^3$ $ha^{-1}\cdot year^{-1}$.

The growth variables evaluated in the Gmelina arborea plantation in Esmeraldas reveal dynamics consistent with the expected growth of this species under tropical conditions, although with nuances that warrant critical analysis.

With respect to diameter at breast height (DBH), the five-year average of 25.3 cm exceeds that reported by Belgis (2022) for Santo Domingo (22.49 cm $ha^{-1}\cdot year^{-1}$) and is slightly lower than in Valencia (27.15 cm $ha^{-1}\cdot year^{-1}$), considering a one-year temporal difference. Similarly, Saltos (2022) reported an average of 20.1 cm $ha^{-1}\cdot year^{-1}$ in Pangua, lower than that of Esmeraldas.

DBH growth is influenced by edaphic and climatic factors, as widely documented. Regarding total height, the MAI in Esmeraldas was 3.61 $m\cdot year^{-1}$, considerably lower than in Santo Domingo (5.91 $m\cdot year^{-1}$) and Valencia (5.52 $m\cdot year^{-1}$). This reflects local edaphic and water limitations.

The commercial volume achieved was 40.89 $m^3$ $ha^{-1}\cdot year^{-1}$ and total volume 45.10 $m^3$ $ha^{-1}\cdot year^{-1}$, figures lower than those observed in plantations located in regions with higher rainfall. While satisfactory, these results suggest the need for intensive plantation management and the promotion of adapted genetic material.

Silvicultural planning should also consider future climate projections to ensure plantation viability. The vulnerability of tropical carbon sinks to climate change makes it imperative to strengthen resilient management strategies, applying landscape approaches to reconcile forest productivity and ecological conservation.

In this context, well-managed secondary forests and plantations constitute critical reserves of ecosystem resilience. Nevertheless, future scenarios also predict increased biotic threats, such as the proliferation of forest pests, reinforcing the need for ongoing monitoring and adaptive management strategies.

Ultimately, the strategic establishment of plantations can contribute not only to sustainable timber production but also to the ecological restoration of degraded landscapes.

## CONCLUSIONS

This study confirms that the growth of *Gmelina arborea* under the edaphoclimatic conditions of Esmeraldas exhibits notable development rates, although these are lower compared to the sites of Santo Domingo and Pangua. These results underscore the importance of edaphic, climatic, and silvicultural variables in determining forest yield. The variability in growth parameters—such as diameter at breast height, total height, and commercial volume—demonstrates that optimizing silvicultural management and implementing strategic interventions are essential to maximize productivity and ensure plantation sustainability.

The data obtained, with an average commercial volume of 45.10 $m^3$ $ha^{-1}$ $year^{-1}$ and a mean annual diameter increment of 5.06 cm, reflect the potential of *Gmelina arborea* as a fast-growing species with high economic profitability, provided that optimal conditions are met and management practices are adapted to local particularities. The disparity in growth rates observed at different sites highlights the need to design conservative and adaptive silvicultural intervention schemes, taking into account specific edaphoclimatic characteristics to promote survival and efficient growth.

Furthermore, the results emphasize the relevance of continuously improving growth prediction models by incorporating environmental variables and potential climate changes, with the aim of optimizing sustainable forest management strategies. It is imperative to strengthen silvicultural actions to improve the survival and productivity of *Gmelina arborea*, thereby contributing significantly to forest sovereignty and regional economic development through the rational and sustainable use of this fast-growing species.

## ACKNOWLEDGMENTS

To the Forest Engineering Program (CIF) at Universidad Técnica del Norte (UTN), for the support and facilities provided for the use of the Xylotheque Laboratory, as well as the equipment and machinery of the Wood Innovation Center during the execution of this research in Ibarra, Ecuador.

Special thanks to the Herbarium of Universidad Técnica del Norte (HUTN), with an ANNUAL OPERATING LICENSE FOR EX SITU CONSERVATION AND MANAGEMENT FACILITIES, code Patent No. MAATE-MCMEVS-2023-035. This institution holds certification and coding for: 1) AUTHORIZATION FOR THE COLLECTION OF SPECIMENS OF BIOLOGICAL DIVERSITY SPECIES; 2) CODE

MAATE-ARSFC-2023-0036, which accredits Eng. Hugo Paredes Rodríguez as the responsible technical researcher.